\documentclass[
aps, reprint,
runinaddress,
 amsmath,amssymb,
 aps, physrev,
]{revtex4-2}

\usepackage{graphicx}
\usepackage{dcolumn}
\usepackage{bm}
\usepackage{subcaption}
\usepackage[export]{adjustbox}
\usepackage[T1]{fontenc}

\begin{document}

\preprint{APS/123-QED}

\title{\textbf{Super-molasses returns: All optical near-resonance laser cooling and trapping of neutral atoms from background vapor.}
}%

\author{Matt Himsworth}
\author{Chester Camm}
\author{Max Carey} %
\author{Jack Saywell}
\author{Jonathan Woods}%
\author{Vilius Atko\v{c}ius}%
\author{Florence Concepcion}%
\author{Konstantinos Karakostas}%
\author{Hannah Brady}%
\author{Doruk Tan Atila}%
\author{Ellie Heywood}%
\author{Alex Jantzen}%
\author{Andrei Dragomir}
\affiliation{Aquark Technologies Ltd, Eastleigh, S050 4SR, UK}
\email{Contact author: m.himsworth@aqrkt.com}

\author{Christopher Morley}%
\affiliation{School of Physics \& Astronomy, Nottingham University, NG7 2RD, UK}

\author{James Bateman}%
\affiliation{Department of Physics, Swansea University, SA2 8PP, UK}

\date{\today}

\begin{abstract}
Laser cooled and trapped atoms have been the workhorse of atomic physics for the past four decades. The predominant method has been the highly versatile Magneto-Optical Trap. We describe an alternative laser trap involving a simple geometry of collimated laser beams that provides both a velocity and position dependent restoring force such that a dense cloud of cold atoms is formed. This technique produces similar atom number ($>10^6$) and density ($10^{10}$\,atoms/cm$^{3}$) to the Magneto-Optical Trap, albeit with \emph{no magnetic field}. The beam geometry is compatible with conventional sub-Doppler cooling techniques, allowing the trapped cloud to be cooled to $< 10~\mu$K. We demonstrate the validity and robustness of the trap by capturing $^{87}$Rb atoms directly from the background vapor and provide a theoretical discussion of the underlying principles. This trap has many unique properties that make it highly suitable for quantum sensing, timing, and computing applications as well as a new tool in fundamental science and metrology.

\end{abstract}

\maketitle


\section{\label{background}Background\protect}

The ability to confine, cool and manipulate atoms has unquestionably been a turning point in our understanding of atomic, molecular, and quantum physics since its invention nearly half a century ago. The magneto optical trap (MOT) has been the workhorse behind this capability and continues to be a key tool for fundamental physics through to applied quantum technologies in sensing, timing and computation.\par

During the early development of laser cooling and trapping in the late 1980s many research groups measured temperatures of the trapped atoms far below the limit imposed by simple Doppler cooling models at the time~\cite{lett1988observation}. In the following year two groups on either side of the Atlantic independently submitted theoretical explanations that identified Polarization Gradient Cooling (PGC, also known as Sisyphus cooling) as the mechanism behind the enhanced cooling effect ~\cite{ungar1989optical, dalibard1989BelowDoppler}. These theories relied on the internal structure of the atoms to scatter additional energy via the polarization-dependent modulation of the energy of the atomic state.\par

\begin{figure}
    \centering
    \includegraphics[width=1\linewidth]{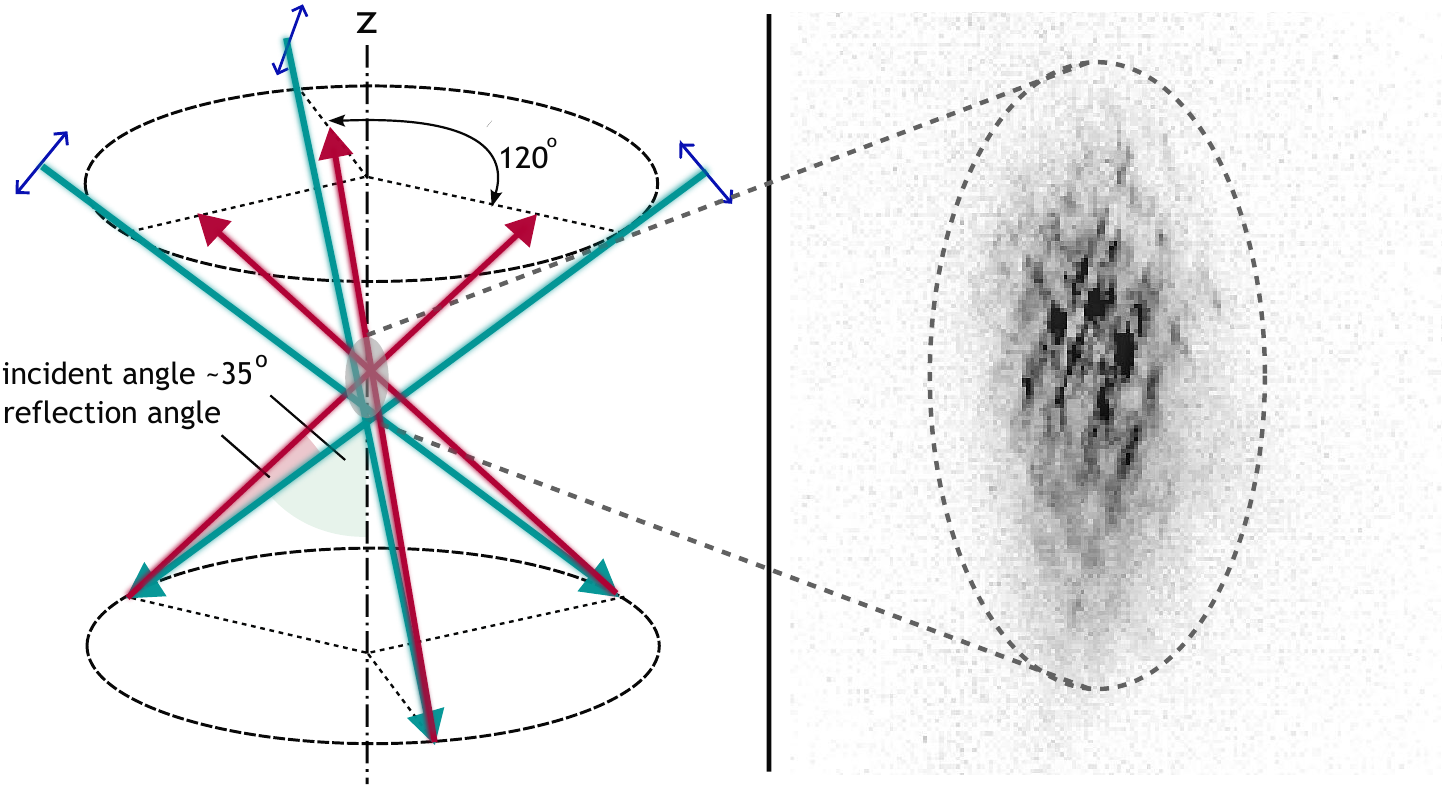}
    \caption{\textbf{Left:} The laser beam geometry used in this paper to produce the super-molasses trap, with incident (green) beams forming a tripod and angled from the vertical by $35^{\circ}$ and retro-reflected (red) beams misaligned by very small angles. Polarization is linear as shown in blue, although the trap can operate with circular and combinations of polarization. \textbf{Right:} A (inverted greyscale) fluorescence image of the trapped ensemble of rubidium atoms. The laser beams have been aligned to produce an extended cloud to highlight the lattice structure. The vertical dimension of the cloud is approximately 3\,mm. Denser and more spherical clouds can be obtained with different beam reflection angles.}
    \label{fig:lattice}
\end{figure}

However, another observation - made during this same period of discovery - has seen very little exploration since. It was noted by Chu et al~\cite{chu1987laser} that when the MOT counter-propagating laser beams were slightly misaligned the number of atoms collected inside the molasses increased significantly, by well over an order of magnitude, and remained trapped for several seconds. This phenomenon was referred to as ‘super-molasses’  to distinguish it from the treacle-like slowing force - optical molasses - of Doppler cooling. Several partial explanations of super-molasses have been proposed including `race track modes’~\cite{ walker1990collective, bagnato1994dynamical}, the enhanced damping of trapped atoms through combinations of optical standing and traveling waves~\cite{ bigelow1990decreased}, and the channeling of atoms in interference patterns~\cite{bigelow1990observation}. Despite these attempts, a complete explanation of the super-molasses phenomenon remains elusive.

Several papers have noted related effects in which unexpected cooling and trapping of atoms occurs in the absence of a magnetic field gradient, for example  by Hope et al.~\cite{hope1994optical} and recently Sharma et al.~\cite{sharma2018all}, with no complete theoretical description. Both traps involve counter-propagating beams along the 3 orthogonal axes, as is typical for MOTs, and were an unexpected discovery. The former case required linearly polarized light, a very good vacuum, and large diameter ($\geq$10\,mm) laser beams which are slightly (0.5-1$^{\circ}$) misaligned when retro-reflected to produce the brightest cloud of trapped atoms collected from background vapor. The suspected mechanism for trapping was the dipole force, and the atom number and density were approximately an order of magnitude lower than the MOT. The latter case also involved slightly misaligned beams and similar centimeter diameter beams but operated with both linear and circular polarization and had to be loaded from a standard MOT before the magnetic field was extinguished. The authors also noted that the temperature of the cloud was not uniform along each axis and that the zero-magnetic field trap worked under various experimental conditions (polarization, intensity, alignment, etc) albeit only within a narrow region of each combination. The authors consider a number of theoretical processes that could result in such a trap and suggest that it may be related to a super-lattice dipole trap.\par 

In this article, we describe a simple laser beam geometry which uses the collimated optical fields typically used in a MOT, but one that both efficiently cools and confines the atoms without the need for a magnetic field. The trap loads quickly and cools directly from the background vapor. It produces a dense cloud of atoms at sub-Doppler temperatures, albeit with a non-uniform density as expected for a dissipative optical lattice (Figure \ref{fig:lattice}). It is our belief that the process is the same mechanism that produced the original super-molasses observation, as well as subsequent findings, so we shall refer to it as a `Super-Molasses Trap' (SMT). We suggest a potential physical description of the effect based on enhanced sub-Doppler cooling mechanisms that provide sufficient cooling to allow the atoms to be confined within a near-resonant dipole trap formed by interference between the laser beams.

\section{\label{results}Results\protect}
 We demonstrate the SMT by cooling and trapping $^{87}$Rb atoms from a background vapor and expect the method is applicable to most atomic species that can be laser-cooled and have non-zero nuclear spin so that PGC forces can occur. The key difference from a MOT is the incident and reflected beam angles and, to some extent, their polarization. A description of the experimental setup is provided in Appendix \ref{setup} and is identical to most MOT setups apart from the laser beam geometry and lack of magnetic field coils (we do retain bias field coils to null residual magnetic fields to obtain the lowest temperatures). \par
 In the following data, three collimated incident beams form a tripod configuration with an angle of $\sim35^\circ$ from the vertical $z$-axis (Figure \ref{fig:lattice}) and are retro-reflected. The exact incident angle is not crucial and we have demonstrated efficient trapping between 30-40$^\circ$, however 35.3$^\circ$ is the superlattice optimum ~\cite{grynberg2001cold} and also produces the most spherical and uniform trapped ensemble. The incident  and reflected beams have linear polarization with P orientation, similar to the superlattice trap geometry described by Leonard and Olsen~\cite{leonard2025atom}.\par

\begin{figure}
    \centering
    \includegraphics[width=0.6\linewidth, angle=90]{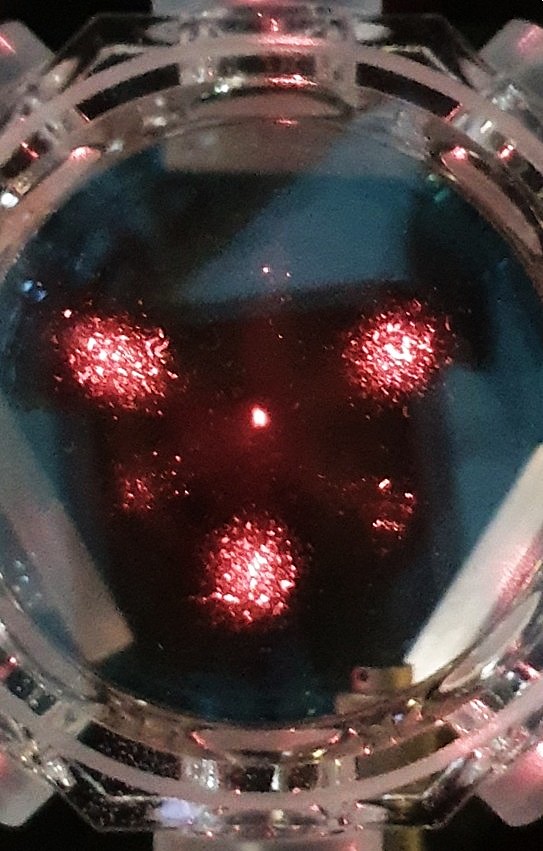}
    \caption{A photograph of the SMT taken along the z-axis of Figure \ref{fig:lattice}). The three retro-reflected beams can be seen scattering on the top chamber window, at the center of which is the brighter cold atom cloud.}
    \label{fig:photo}
\end{figure}
 
 We find the SMT will form  with a variety of polarizations, but the lin-lin-lin geometry appears optimal and is most resilient to magnetic fields (see Appendix \ref{magnetic}). The beams are then retro-reflected with a mirror slightly misaligned from normal, predominantly towards the $z$-axis. The alignment process affects both the cloud shape, density, temperature uniformity, and position of the cloud through interference effects. The power in each incident beam does not need to be equal, nor particularly phase stable; in this data-set the three beams are provided by a polarization maintaining fiber splitter and the whole vacuum chamber and optics are on a non-vibration isolated table. Nevertheless, the position of the cloud is remarkably stable as shown in Appendix \ref{stability}. \par

Figure \ref{fig:num_temp} shows the atom number measured via absorption imaging of the cloud expansion after the beams are extinguished ~\cite{smith2011absorption}. We see that the steady state atom number can reach $>2\times10^{6}$ in a cloud with $\sim 600\,\mu$m diameter. The peak number coincides with a detuning of $-1\,\Gamma$ and an intensity of 22\,mW/cm$^2$ per beam (a saturation parameter of 6-7 on the closed cooling transition. $\Gamma$ is the transition linewidth.) With a comparable beam diameter one might expect a MOT to achieve an order of magnitude more atoms ~\cite{hoth2013atom}, but we find the atom density in the SMT is on the order of $2\times10^{10}$/cm$^{3}$ which is close to a radiation pressure limited MOT. This is likely due to the interference fringe volume required to achieve the tightest confined cloud. Finding the global optimum alignment of beams to produce the largest, deepest, trap is non-trivial and there may be better configurations beyond the data presented here.  

\begin{figure}[t]
    \centering
    \includegraphics[width=1\linewidth]{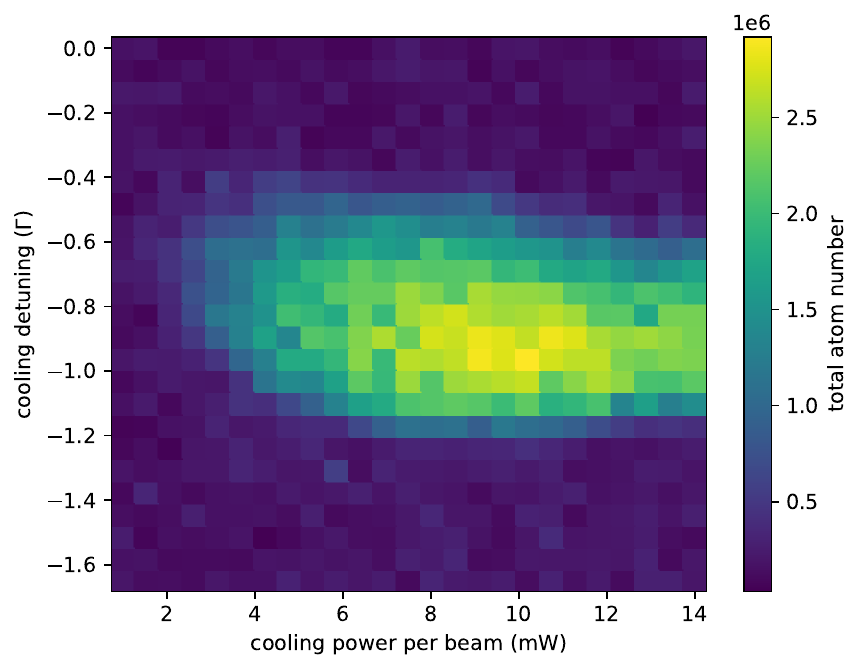}
    \caption{Dependence of atom number in the SMT on laser detuning and power. We see a peak of approximately 2.5 million atoms at -1$\,\Gamma$ detuning and 10\,mW per beam. The peak atom number is highly dependent on beam alignments but once a global maximum is found the atom number is quite consistent for the same beam width.}
    \label{fig:num_temp}
\end{figure}

\begin{figure}[t]
    \centering
    \includegraphics[width=1\linewidth]{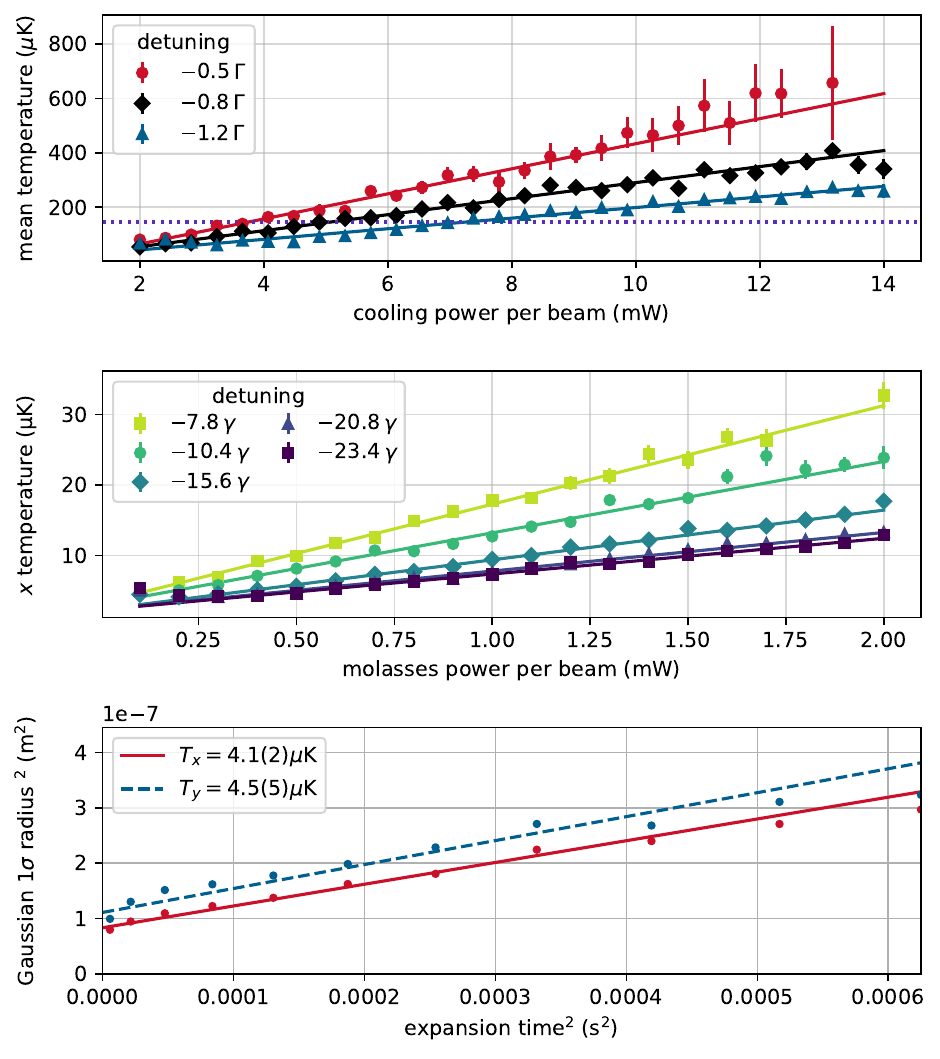}
    \caption{\textbf{Top}: The trapped ensemble temperatures before the `molasses' stage for a selection of detunings. The solid lines are linear fits to guide the eye, and the dashed line indicates the Doppler Temperature for $^{87}$Rb. \textbf{Middle}: The trapped ensemble temperature after a optical molasses stage during which the laser detuning is shifted to the displayed frequency in 2\,ms, whilst the power is linearly ramped down to different final values. Solid lines are linear fits to guide the eye. \textbf{Bottom}: An example of the ensemble radius expansion data (dots) used calculate the temperature (lines). We see the temperature is uniform along both axes.}
    \label{fig:molasses}
\end{figure}

We find a much narrower region in which the SMT will operate compared to a MOT in both laser detuning and power. There is a small dependency in optimum laser detuning with intensity, as expected with light shifts within the dipole trap. The peak atom number is found at -1$\,\Gamma$ detuning, compared to -2.7$\,\Gamma$ for a MOT \cite{lindquist1992experimental}. The ensemble temperatures were uniform along all axes to within $1\ \mu$K when the beams were aligned to generate a nearly spherical cloud and predominantly around or above the Doppler temperature of $146\,\mu$K, as shown in Figure \ref{fig:molasses}. Higher laser power and lower detuning result in higher temperatures. Adding an optical molasses stage in which the cooling beam is momentarily lowered in power and tuned to -23$\,\Gamma$ achieves low $\mu$K temperatures (this is also employed in MOTs to reach equally low temperatures). 

We predict the optical dipole trap depth is three orders of magnitude lower than a MOT (see Section \ref{discussion}) and approximately equal to the ensemble temperature before the molasses cooling stage. Therefore, the SMT will be more sensitive to background gas collisions. The chamber background pressure is $\sim10^{-11}$\,mbar, and we find peak atom number at a Rubidium vapor pressure of mid-to-high $ 10^{-10}$\,mbar as indicated by the ion pump current. Above $\sim5\times 10^{-9}$\,mbar the atom number reduces depending on the laser intensity. A MOT in comparison generally collects more atoms with increasing vapor pressure and still operates into the $10^{-7}$\,mbar region. 

\section{\label{discussion}Discussion\protect}
Like those before us, the discovery of the SMT was serendipitous during exploration of different MOT beam geometries ~\cite{dragomir2018cold}. A full theoretical quantitative description of the cooling and trapping processes is non-trivial and is beyond the scope of this report. Here, we provide potential explanations that may progress understanding. The reason why cooling and trapping of atoms with solely optical fields is unexpected is due to the optical equivalent of the Earnshaw theorem~\cite{ashkin1983stability} that states $\nabla \cdot F_S=0$ where $F_S$ is the scattering force on a particle with scalar polarizability. In essence, the scattering force alone cannot produce a stable restoring force on a dipole in all dimensions because there are no sources or sinks of optical fields in free space. This limitation has been circumvented by manipulating the internal states of the atom via Zeeman shifts (as in a MOT), AC Stark shifts (as in a dipole trap), or spatially-imbalanced optical pumping ~\cite{bouyer1994atom}. \par

One common mechanism suggested for purely optical cooling and trapping is the bichromatic force, or retarded dipole potential ~\cite{wkasik1997non}. The trapping force is a result of two or more optical wavelengths whose nonlinear interference produces a deep rectified dipole potential. We can immediately neglect this mechanism because the trap will operate equally with both co-propagating cooling and repump beams, with an orthogonally polarized repump, and a spatially separated repump beam.\par

The experimental results suggest the overall cooling and trapping mechanism is a dissipative optical lattice produced by the misalignment between beams resulting in interference fringes with a variety of pitch lengths, from $\lambda/2$ to several millimeters~\cite{steane1992radiation}. At specific alignment the interference produces intense global intensity maxima into which the atoms are funneled and ultimately confined at a single point. These large interference structures provide sufficient volume  to trap significant numbers of atoms without excessive collisional loss found in more tightly confined dipole traps ~\cite{schlosser2002collisional,prentiss1988densitylosses}. Moreover, the incident beam angles are close to a superlattice, and as such produce a highly stable configuration that is resilient to fluctuations in individual input beam phase ~\cite{leonard2025atom, schadwinkel1999magneto}. Similar confining effects have been found in 2D misaligned optical beams ~\cite{alfaro2021long} and we also find trapped atoms populate high $m_F$ `stretched' states in the SMT. Figure \ref{fig:lattice} (right) is an image of a trapped cloud where the beams have been misaligned to highlight the extended optical lattice structure. Simulations of the field shows variations in the ellipticity of the trapping light that match the lattice periodicity and scale. Further alignment can confine the atoms into a tight, dense and typically elliptical cloud similar to that found in a MOT, as shown in the photograph in Figure \ref{fig:photo}.\par

The near resonance dipole trap depth  for a collimated beam can be calculated with ~\cite{bjorkholm1978observation}
\begin{equation}
    U=\frac{\hbar\Delta}{2}\ln\left(1+\frac{ I}{ I_{S}}\frac{\Gamma^2/4}{\Delta^2+\Gamma^2/4}\right)
\end{equation}
where $\hbar$ is the reduced Planck's constant, $\Delta$ is the detuning from resonance, $I$ is the beam intensity, $I_S$ is the saturation intensity of the transition. Negative detuning leads to trapping at the intensity maximum. We find that for the parameter space explored here the trap depth is on the order of 400\,$\mu$K for the peak atom number in Figure \ref{fig:molasses} and are consistently 100-200$\,\mu K$ above the cloud temperature before the molasses cooling stage. This is over three orders of magnitude smaller than a MOT trap depth~\cite{haw2012magneto}. This explains the high sensitivity of the trap to collisional losses from the background vapor.\par

The main questions are why this configuration collects so many atoms very quickly, but does not suffer from excessive heating/loss? In a MOT the steady state atom number is a balance between loading rate (determined by trap depth/capture velocity, beam overlap volume, and background vapor pressure) and losses from background gas `hot' collisions, intratrap `cold' collisions, and re-absorption of radiation between atoms. This all applies in the SMT, albeit with a much lower trap depth and higher scattering rate due to smaller detuning from resonance. Indeed, we see trap loss from background vapor collisions higher than a MOT, even at peak loading rates (see Figure \ref{fig:loading_v_pressure}). There may also exist an additional heating term from atoms excited to dipole repulsive states ~\cite{dalibard1985dressed}. Clearly, the loss elements should be equal or greater than a MOT and hence why near resonant dipole traps historically appeared impractical. We must conclude that for the SMT to be so efficient, the loading rate must be significantly enhanced. This is either through accessing higher velocity atoms within the thermal background, greater optical forces, or increasing the effective trap volume~\cite{haw2012magneto}.\par   

\begin{figure}
    \centering
    \includegraphics[width=1\linewidth]{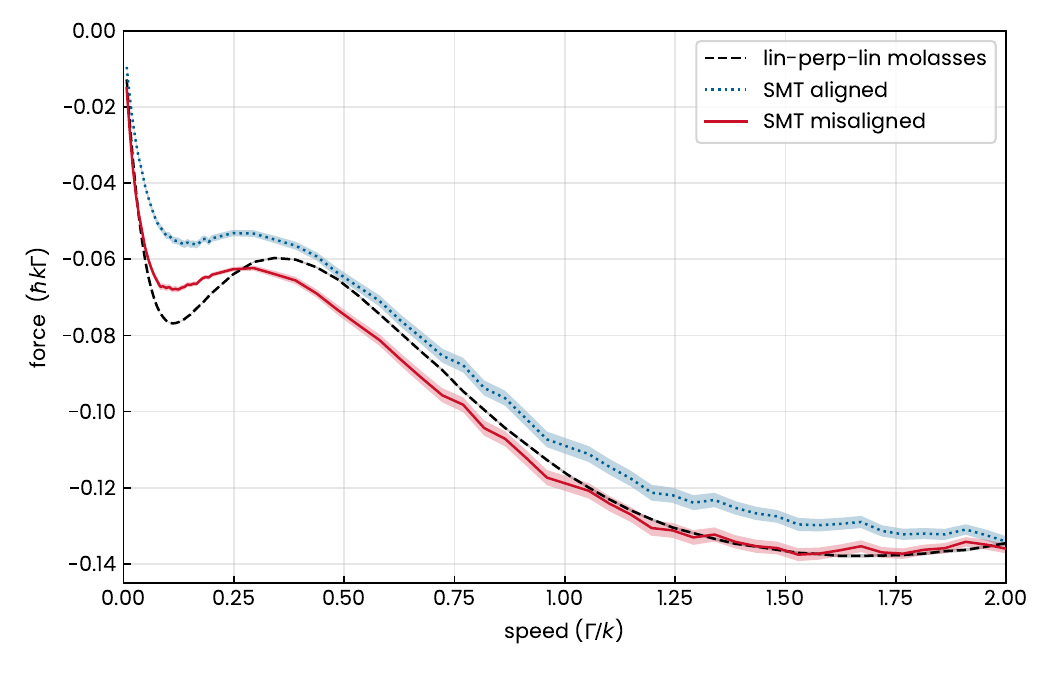}
    \caption{The total optical force - simulated using the approach of Delvin and Tarbutt~\cite{devlin2016three} - averaged over an isotropic distribution of 2000 random directions and initial positions as the atom is dragged through the SMT optical field with different speeds. Negative values signify cooling. The force is shown for the SMT beam geometry with all the retroreflected beams perfectly aligned (blue) and with one beam misaligned by 2$^{\circ}$ (red). We compare against orthogonally aligned lin-perp-lin beams found in a typical optical molasses geometry in a MOT (black dashed). Laser powers and detuning are the same in all plots. Misaligning the beam enhances the cooling force across speeds below $2\,\Gamma/k$ in the SMT geometry, and also shows an improvement over lin-perp-lin optical molasses in the range 0.25-1.0\,$\Gamma/k$. The enhancement is a result of interference between the beams that creates long-pitch polarization gradients. In this model we neglect the Gaussian beam shape. The shaded region indicates a 4-$\sigma$ confidence region determined through bootstrapping~\cite{efron1994introduction}.}
    \label{fig:theory}
\end{figure}

 PGC is produced via optical pumping between AC-Stark shifted atomic states which causes the atom to lose additional energy as it passes through optical standing waves with polarization gradients. Typically, PGC occurs only at very low velocities as the force is dependent on the time for atoms to travel across a single fringe of the standing wave and the optical pumping time. In the SMT, there are several different pitch scales alongside very complex polarization structures. This may allow PGC to extend to larger velocity classes and a much greater number of atoms in the thermal background. According to the Reif model~\cite{haw2012magneto}, the number of atoms cooled is proportional to the fourth power of their capture velocity, so even a small extension of the PGC force to higher velocities can have a significant effect in trap loading. This is part of why the optimal laser detuning in a MOT is greater than the theoretical prediction from the simple Doppler cooling model. Moreover, PGC forces scale as $k^2$ as opposed to Doppler cooling scaling which scales linearly with the wavevector $k$; so can be significantly stronger than purely scattering forces ~\cite{dalibard1989BelowDoppler}. We have modeled this effect following the approach of Delvin and Tarbutt \cite{devlin2016three} and found there is a significant enhancement of the cooling force when one beam is misaligned as shown in Figure \ref{fig:theory}.

Another potential mechanism is enhanced scattering of the atomic velocity through the optical potential landscape. Instead of a nearly linear trajectory toward a singular restoring point in a MOT, the atoms in an SMT experience Brownian motion or L\'evy flight statistics~\cite{afek2023colloquium}, which could significantly increase time spent in the beams, thus providing an effective enhancement of the trapping volume. At low velocities this is certainly the case and has a significant effect on diffusion mechanisms ~\cite{lett1989optical, steane1992radiation}. Similar mechanisms have been suggested and demonstrated for 3D cooling and trapping in a quasi-periodic speckle field~\cite{horak1998atom, boiron1999trapping}. One may expect two or more components in the Maxwell-Boltzmann distributions of velocities during cloud expansion. This requires further exploration, but such features have not been observed in our initial measurements, potentially due to the mixing of several components. \par 

The difficulty in modeling the SMT is due to the importance and complexity of spontaneous and stimulated relaxation processes near resonance and within a spatial varying optical potential with multiple scale factors. It is expected a full dressed state model is more appropriate, but difficult to model numerically and challenging to reduce down to the key physical mechanisms ~\cite{dalibard1985dressed}. We hope the relative ease of setting up the SMT will provide more opportunities for the community to explore these mechanisms.

\section{\label{summary}Summary and outlook\protect}

We have demonstrated a highly efficient dissipative optical lattice, or near resonant dipole trap, capable of cooling and trapping atoms from the background vapor, with performance equivalent to a MOT but with no requirement for a magnetic field. This type of trap may provide a number of unique capabilities that will significantly benefit atomic physics, and quantum technologies. Most notable is the ability to trap multiple clouds in close proximity as a result of the lack of magnetic fields. This property may also benefit mixed-species traps and potentially reduce the Dick effect in precision measurements. The lack of focused beams means that the optical lattice can have very large dimensions, which could benefit neutral atom quantum computing and sensing of field gradients. Momentarily detuning the beams further from resonance can achieve even lower temperatures ~\cite{leonard2025atom}, lower scattering and allow for efficient all-optical generation of Bose-Einstein condensates. The high stability of the trap is important for atom interferometry, in particular gradiometry, where initial positions of the cloud affect resolution. The ability to quickly load the trap from very low background pressures will improve signal to noise and collisional effects. Although we have discussed several potential qualitative explanations of the cooling and trapping mechanisms, a full quantitative explanation remains elusive but initial results are encouraging. Furthermore, the relative ease of creating a super-molasses trap with a large number of trapped atoms should open experimental pathways to improve our understanding of this 40-year old mystery.

\begin{acknowledgments}
We wish to acknowledge the support of Tim Freegarde, Peter Horak and Elliot Bentine. 
\par
this work was supported by the Defence Science and Technology Laboratory (Dstl) for the PhD studentship of Andrei Dragomir, the UKRI grant EP/R041806/1, and Innovate UK grants 10028190 and 10032699. The geometry of the SMT laser beams is covered by patents: EP4345846A2, US11763956 and US12080442. 

\end{acknowledgments}

\appendix
\section{\label{setup}Experimental setup}

Our laser cooling system consists of two distributed Bragg reflector (DBR) diode lasers (Photodigm Iso-Bragg) - cooling and repump which are phase locked to a single reference DBR laser which is stabilized using saturated absorption spectroscopy to the $^{85}$Rb $F=3\to F'=4-2$ crossover ~\cite{himsworth2010rubidium}. The imaging laser is an ECDL, also phase locked to the reference laser, and tuned on-resonance to the $^{87}$Rb $F=2\to F'=3$ cooling transition. The cooling laser is tuned from this transition via the values provided in the relevant figures, and we tune the repump laser to the $^{87}$Rb $F=1\to F'=2$ transition. The cooling laser is amplified with a Thorlabs BOA780P semiconductor optical amplifier. Acousto-Optic modulators provide laser power control and switching. The cooling and repump lasers are combined and then split by a Phoenix Photonics $3\times3$ fibre splitter. The outputs are all collimated to $7.5\,$mm  ($1/e^2$) diameter beams. The cooling and repump beams have linear polarization as shown in Figure \ref{fig:lattice}. The rubidium source is a SAES alkali metal dispenser, and the vacuum system comprises of a bespoke all-glass octagonal cell from Precision Glass Blowing and a 20lt/s ion pump from Gamma vacuum.\par  

 The optimal misalignment requires careful iteration but in general a compact cloud can be found quickly by first perfectly retro-reflecting all three beams, then misaligning one towards the z-axis by approximately half a beam-width at the position of the atom cloud (see Figure \ref{fig:lattice}). This is followed by smaller misalignment of the remaining two beams, and then further iterative adjustment of all beams until the cloud is in the correct position and size. Offset angles of a fraction of a degree can result in fringe pitches of hundreds of microns - the same dimensions as the most compact cloud. Initially a diffuse and extended cloud will appear and adjustment of all the mirrors in turn can confine the atoms in a single dense cloud, an extended lattice or even a long thin cloud.\par

The cloud properties are determined through absorption imaging, with a beam of $11\,$mm ($1/e^2$) diameter, and $10\,\mu$W of power. This is aligned orthogonal to the z-axis, through the atom cloud, and onto a camera lens system which is focused onto the atom cloud. Three images are taken to provide background subtraction: the atom cloud present, without the atom cloud present, and no incident light ~\cite{smith2011absorption}. The normalized images of the expanding cloud are fitted to 2D Gaussian functions from which the width and center position are determined.

\section{\label{stability}Trap stability}

\begin{figure}[t]
\centering
  \includegraphics[width=1\linewidth]{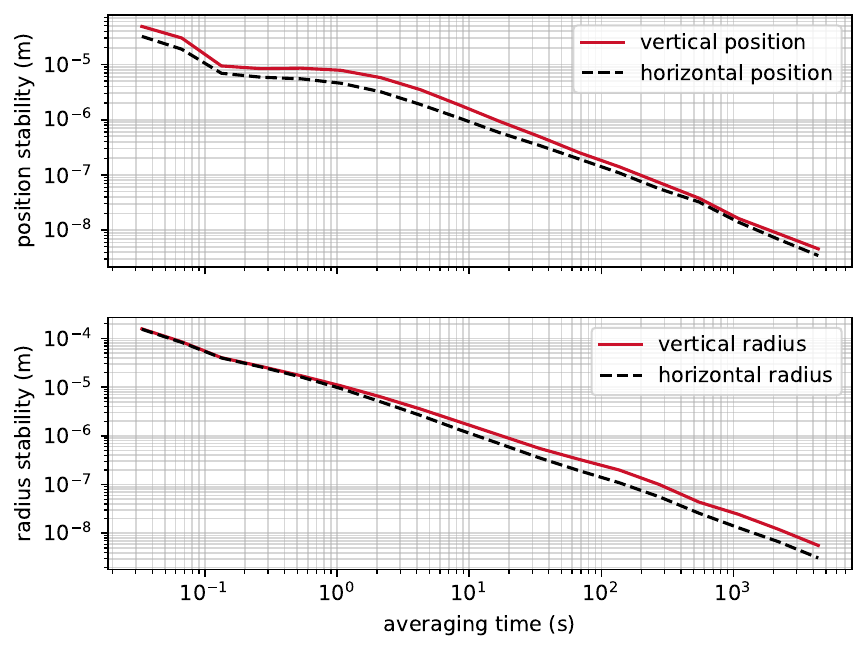} 
  \caption{The Allan Deviation of the trap centre (\textbf{Top}) and size (\textbf{Bottom}) over several timescales. The center is determined by fitting of a 2D Gaussian to absorption image. Short term instability is dominated by vibrations of the undamped table due to nearby traffic.}\label{fig:stability}
\end{figure}

The positional stability of the trap (shown in Figure~\ref{fig:stability}) is remarkably high and far exceeds that of a conventional 6-beam MOT (single beam pyramid and gratings MOTs have significantly better stability than typical MOTs). The vacuum chamber and optics are mounted on a worktop with no vibration damping, but the SMT possesses nano-meter stability over several minutes. This is due to the trap position being dependent solely on beam alignment and optic mount rigidity. This is compared to a MOT where the trap center depends additionally on a complex interplay between magnetic fields, beam intensity balance and laser detuning.

\section{\label{density}Trap density}

\begin{figure}[t]
    \centering
    \includegraphics[width=1\linewidth]{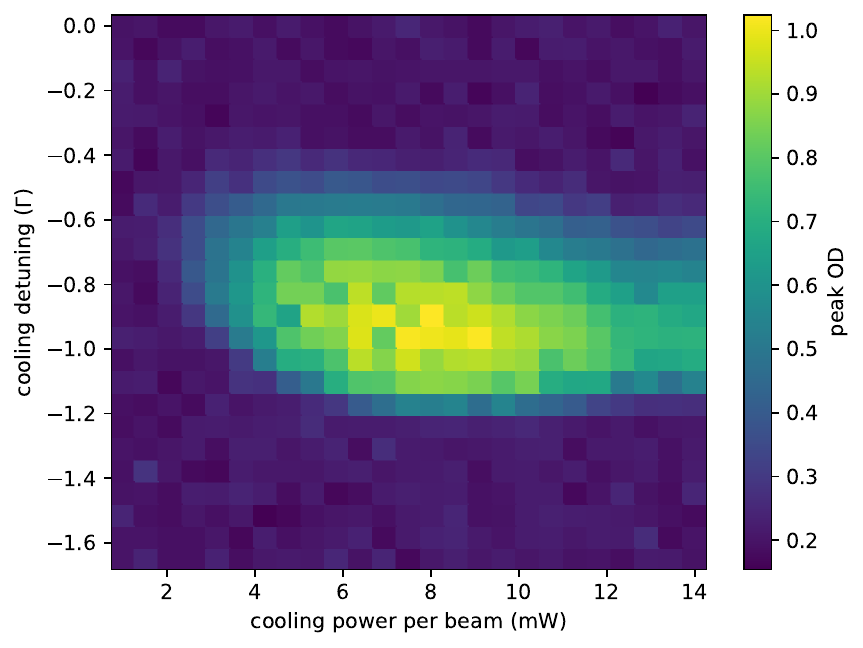}
    \caption{Peak resonant optical density of the SMT as a function of power in each beam and also detunings of the cooling beams from resonance. Maximum values are of order unity, corresponding to a peak number density on the order of $10^{10}$ atoms/cm$^{3}$. }
    \label{fig:desnity}
\end{figure}

The lattice structure of the SMT results in a non-uniform cloud shape. Careful manipulation of the beams can result in a singular `global maximum' intensity fringe such that all the atoms are confined to a quasi-single point. We have measured the maximum optical density of the cloud in this configuration for a variety of trap parameters as shown in Figure \ref{fig:desnity}. Optical density follows a similar trend as atom number, as expected, with peak values of unity, corresponding to a peak number density but at a slightly lower cooling beam power.

\section{\label{loading}Trap loading rate}

\begin{figure}[t]
  \centering
  \includegraphics[width=1\linewidth]{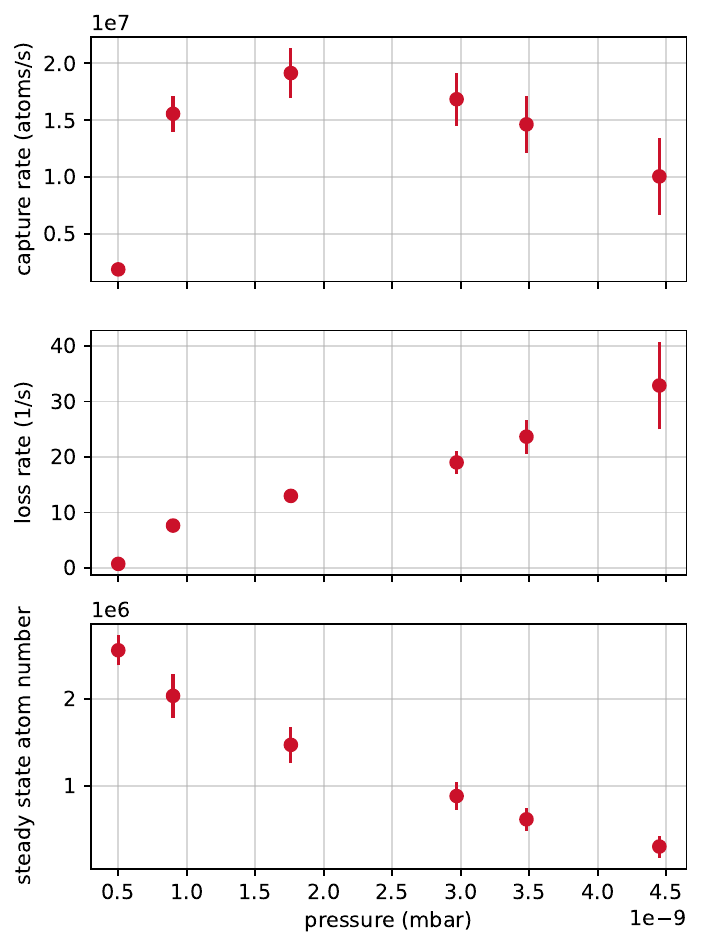}
\caption{Measured loading rates, loss rates, and steady state atom numbers for the SMT as the rubidium pressure is varied from $5-50\times10^{-10}$~mbar (inferred from ion pump current). We observe a peak loading rate at a pressure of approximately 2$\times10^{-9}$ mbar but with a compromise of half the peak atom number.} 
\label{fig:loading_v_pressure}
\end{figure}

 For a MOT, the loading rate is proportional to the square of trap depth, and thus laser intensity and detuning and magnetic field gradient. However, we find the trap loading time constant of $\sim0.5$\,s is relatively uniform across the range of intensities and detunings explored. Whereas MOTs exhibit more trapped atoms and faster loading rates with increased vapor pressure, the SMT operates most efficiently at very low vapor pressure. We have reached loading rates of $2\times10^7$ atoms/s at $2\times10^{-9}$\,mbar, albeit with the steady-state atom number halved, as shown in Figure~\ref{fig:loading_v_pressure}. In our lowest vacuum chambers (background, non-Rb, pressure) with $< 10^{-11}$\,mbar, we can still find a cloud several tens of minutes later in the trap after the dispensers have been extinguished, albeit with significantly lower atom number. This is partially due to low background collisions but mainly efficient loading from dilute residual rubidium vapor.

\section{\label{magnetic}Magnetic field sensitivity}

For applications outside the laboratory, it is advantageous to be insensitive to residual magnetic fields, while within the lab applying a uniform magnetic field to provide a quantization axis is often necessary. We characterized the sensitivity of the SMT by measuring the atom number as uniform external fields were applied along the vertical (z) axis and along an orthogonal axis, with the cooling laser detuning fixed at $-1\Gamma$. The trap shows little variation up to 1\,Gauss, with less than a 10\% reduction in atom number over this range, and the response is symmetric about zero field on both axes. Higher fields may be achievable via detuning of the cooling beams to match the Zeeman shift ~\cite{sharma2018all}. Unlike a MOT, external fields do not shift the position of the cloud and just result in lower density around the same central point.
We note that the magnetic field does affect the cloud temperature, and the low temperatures shown in Figure~\ref{fig:molasses} are in a magnetically-nulled environment, similar to a MOT. We find the Lin-Lin-Lin polarization configuration is much less sensitive than circ-circ-circ. Similar insensitivity was noted by Hope et al ~\cite{hope1994optical}.

\bibliographystyle{apsrev4-2}
\bibliography{SMT_2026_refs}

\end{document}